\documentclass[twocolumn,aps,superscriptaddress]{revtex4-1}
\usepackage{palatino}
\usepackage[latin1]{inputenc}
\usepackage{epsf}
\usepackage{amsmath,amssymb}
\usepackage{latexsym}
\usepackage{calc}
\usepackage{color}
\usepackage{shadow}
\usepackage{epsfig}
\usepackage[usenames,dvipsnames]{xcolor}
\usepackage{subfig}

\def\bea{\begin{eqnarray}}
\def\eea{\end{eqnarray}}
\def\ben{\begin{equation}}
\def\een{\end{equation}}
\def\benu{\begin{enumerate}}
\def\enu{\end{enumerate}}

\def\n{n}

\def\sss{\scriptscriptstyle\rm}

\def\1var{(\bx_1...\bx\N)}

\def\sss{\scriptscriptstyle\rm}
\def\x{_{\sss X}}
\def\c{_{\sss C}}
\def\s{_{\sss s}}
\def\xc{_{\sss XC}}

\def\H{_{\sss H}}

\def\S{^{\sss S}}

\def\br{{\bf r}}

\def\bx{{\br t}}

\def\bj{{\bf j}}

\def\bb{{\bf b}}

 \def\n{n}
 \def\sss{\scriptscriptstyle\rm}
 \def\1var{(\bx_1...\bx\N)}
 \def\x{_{\sss X}}
 \def\c{_{\sss C}}
 \def\s{_{\sss S}}
 \def\xc{_{\sss XC}}

 \def\N{_{\sss N}}
 \def\H{_{\sss H}}
\def\ALDA{^{\rm ALDA}}

\begin{document}
\title{Exploring non-adiabatic approximations to the exchange-correlation functional of TDDFT}
\author{Johanna I. Fuks}
\affiliation{Departamento  de  F\'isica  and  IFIBA,  FCEN, Universidad  de  Buenos  Aires,  Ciudad  Universitaria, Pab.  I,  C1428EHA  Ciudad  de  Buenos  Aires,  Argentina}
\author{Lionel Lacombe}
\affiliation{Department of Physics and Astronomy, Hunter College and the Graduate Center of the City University of New York, 695 Park Avenue, New York, New York 10065, USA}
\author{S{\o}ren E. B. Nielsen}
\affiliation{Max Planck Institute for the Structure and Dynamics of Matter and Center for Free-Electron Laser Science, Luruper Chaussee 149, 22761 Hamburg, Germany}
\author{Neepa T. Maitra}
\affiliation{Department of Physics and Astronomy, Hunter College and the Graduate Center of the City University of New York, 695 Park Avenue, New York, New York 10065, USA}

\date{\today}
\begin{abstract}
A decomposition of the exact exchange-correlation potential of time-dependent density functional theory into an interaction component and a kinetic component offers a new starting point for non-adiabatic approximations. 
The components are expressed in terms of the exchange-correlation hole and the difference between the one-body density matrix of the interacting and Kohn-Sham systems, which must be approximated in terms 
of quantities accessible from the Kohn-Sham evolution.  We explore several preliminary approximations, evaluate their fulfillment of known exact conditions, and test their performance on simple model systems for which available 
exact solutions indicate the significance of going beyond the adiabatic approximation.  \end{abstract}

\maketitle
\section{Introduction}

Much effort has been put into the development of more accurate and reliable approximations to the ground state (gs) exchange-correlation (xc)
functional of Density Functional Theory (DFT) \cite{YLT16,B12,B14}.
In order to make use of these approximations in the simulation of dynamical processes one invokes the adiabatic 
approximation: one utilizes a gs xc functional in place of the time-dependent one, and so implicitly assumes the many-body effects in the system are those of a ground-state even if the system is itself not near a ground state. Memory effects arising from the dependence of the xc functional on the history of the density, and on the initial states, are completely absent in these approximations. 
Adiabatic Time-dependent Density Functional Theory (TDDFT) is widely and successfully used to simulate excitation spectra and response properties in finite systems and solids that are initially in the 
gs~\cite{RG84,M16,TDDFTbook12,Carstenbook} and probe the perturbative regime, where the system is only slightly disturbed from its ground state. 
The adiabatic approximation becomes less  justified as we 
move away from the gs, as is the case when the perturbation is strong or resonant or when we start the 
simulation in an excited or arbitrary superposition state. 
 In such situations, which have become of increasing interest in the past decade due to the development of experimental techniques and especially the growing interest in attosecond physics (e.g. Ref.~\cite{NDCPM17}), the adiabatic approximation is at least questionable. 
How severely the adiabatic propagation deviates from the exact dynamics will depend on the system and the particular dynamics, and also on what the observables of interest are. 
For example, highly non-perturbative processes such as resonant dynamics and long-range charge transfer have been shown to be particularly challenging 
for adiabatic TDDFT \cite{RB09, FHTR11,RN11,RN12,RN12b,RN12c,FERM13, FFTAKR13,F16,M17,PI16,HTPI14, OMKPRV15,GBCWR13} and even in the case of field-free evolution of a non-stationary state 
the exact xc functional develops non-trivial features that influence the dynamics but that adiabatic functionals cannot capture \cite{EFRM12,LFSEM14,FNRM16}. On the other hand, for example, the adiabatic approximations give good predictions for photo-emission spectra and photo-angular distributions in clusters~\cite{WDRS15}. 
Given the lack of alternative methods to simulate electron-dynamics in medium to large systems and 
the expanding range of time- and space-resolved experiments involving non-perturbative dynamics 
it is imperative to explore the capability of TDDFT beyond the adiabatic approximation. There have been only a few proposals for  
non-adiabatic xc functionals~\cite{VK96,VUC97,KB06}, but they are not commonly used for a variety of reasons including that they spuriously damp finite systems. Functionals that are explicit orbital-functionals also incorporate memory (e.g. time-dependent exact exchange~\cite{UGG95,LHRC17}) but the orbital-dependence of these commonly only involves exchange which is inadequate to capture the most significant non-adiabatic effects.

In this work we present a new starting point for developing non-adiabatic approximations 
based on an exact decomposition of the xc functional.  The exact decomposition, which is reviewed in section \ref{sec:decomp}, expresses the exact xc potential,  in terms of an interaction component that depends on the xc hole of the interacting system, and a kinetic component that depends on the difference between the one-body density matrices of the interacting and Kohn-Sham systems~\cite{LFSEM14,L99}. 
The idea is to approximate these terms in terms of quantities that are accessible from the evolving Kohn-Sham system and/or from the initial many-body wavefunction, and some preliminary approximations are proposed 
in sections \ref{sec:approxs}.  In section \ref{sec:excond} we analyze these in terms of the fulfillment of exact conditions and in section \ref{sec:runs} 
we test their performance against the exact solution for some model systems. Finally, in Sec.~\ref{sec:concs} we conclude.

\section{Decomposition of $v\xc$ into interaction and kinetic potentials}
\label{sec:decomp}
Equating the equation of motion for the second time-derivative of the one-body density,  $\ddot{n}(\br, t)$, in the physical system, with that in the  
Kohn Sham (KS) system $\ddot{n\s}(\br, t)$,  yields the following equation~\cite{L99,LFSEM14} for
$v\xc(\br, t)$:
\begin{widetext}
\ben
\nabla\cdot\left(n\nabla v\xc\right) =\nabla \cdot \left[ \frac{1}{4}\left(\nabla' - \nabla\right)\left(\nabla^2 - \nabla'^2\right) \left( \rho_1(\br',\br,t) - \rho_{1,s}(\br',\br,t) \right) \vert_{\br'=\br} + n(\br,t)\int n\xc(\br,\br',t) \nabla w(\vert \br'-\br \vert) d^3r'\right] ,
\label{eq:3Dvxc}
\een
where 
$
\rho_1(\br',\br,t) = N\sum_{\sigma_1..\sigma_N} \!\!\int d^3r_2...d^3r_N \Psi^*(\br'\sigma_1,\br_2\sigma_2...\br_N\sigma_N;t) \Psi(\br\sigma_1,\br_2\sigma_2 \dots \br_N\sigma_N;t)$
is the spin-summed one-body density-matrix (1RDM) of the true system of electrons with two-body interaction potential $w(\vert\br - \br'\vert)$, $\rho_{1,\sss{S}}(\br',\br,t)$ is the 1RDM for the Kohn-Sham system,  and $n\xc(\br',\br,t)$ is the xc hole, defined via the pair density through
$P(\br,\br',t) = N(N-1)\sum_{\sigma_1..\sigma_N}\int \vert \Psi(\br'\sigma_1,\br\sigma_2,\br_3\sigma_3..\br_N\sigma_N; t) \vert^2 d^3r_3..d^3r_N 
 = n(\br,t)\left(n(\br',t) +n\xc(\br,\br',t)\right)$\;. The notation $\nabla$ means gradient with respect to $\br$ while $\nabla'$ means gradient with respect to $\br'$. 
 \end{widetext}
Eq.~(\ref{eq:3Dvxc}), with mixed separated boundary conditions (i.e. involving the potential and its gradient at each boundary point separately) has the form of a nonhomogeneous Sturm-Liouville problem for $v\xc(\br,t)$. 
We can decompose this potential into  an interaction (W) and a kinetic (T) component,
\ben
v\xc(\br,t)=v\xc^W(\br,t) + v\c^T(\br,t)
\label{eq:vxcdecomp}
\een
where the interaction part (W) has an explicit dependence  on the xc hole,
\ben
\nabla {v\xc^W} = \int n\xc(\br,\br',t) \nabla w(\vert \br'-\br \vert) d^3\br',
\label{eq:vxcW}
\een
and the kinetic (T) component,
\bea
\nonumber
\nabla v\c^T &=& \frac{\left(\nabla' - \nabla \right)\left(\nabla^2- \nabla'^2\right)\Delta\rho_{1}(\br',\br,t)|_{\br'=\br}}{4n(\br,t)} \\
&+& \frac{1}{n(\br,t)}\nabla \times a(\br,t)
\label{eq:vcT}
\eea
has an explicit dependence on spatial gradients of  $\Delta \rho_1(\br',\br,t) = \rho_1(\br',\br,t) 
- \rho_{1,S}(\br',\br,t)$
evaluated at $\br=\br'$. In Eq.~(\ref{eq:vxcW}) we have assumed that the  right-hand-side approaches zero as $\br\to\infty$, so that no additional term is needed to fulfill the boundary condition. However,  such a term, $\frac{1}{n(\br,t)}\nabla \times a(\br,t)$,  is in general needed in Eq.~(\ref{eq:vcT}) to fulfill the Sturm-Liouville boundary condition; we will show in Appendix~\ref{sec:GTIproof} that without this term, $v\xc$ would not be assured to satisfy the generalized translational invariance condition, which it must. There are cases in which this term is zero, including for finite systems in one-dimension; in many cases we will assume this is the case.

An analogous decomposition of the exact ground-state xc potential~\cite{BBS89,GLB94,GLB96,GB96} has provided insight into features of the gs xc potential, while Ref.~\cite{LFSEM14} explored how non-adiabatic features of the time-dependent xc potential manifest themselves in each of these terms. For example, it was found that the dynamical step features found in Refs.~\cite{RG12,EFRM12,FERM13} predominantly appear in $v\c^T$, while $v\xc^W$ appears generally much smoother.  

In this work we take the first steps to move the decomposition beyond being simply a tool to analyze the exact potential to being a foundation on which to build non-adiabatic functional approximations. 
By virtue of the Runge-Gross theorem \cite{RG84} the exact xc potential is a functional of the density $n(\br,t)$, the initial
many-body state $\Psi(\br_1\sigma_1...\br_N\sigma_N; 0)$, and the initial Kohn-Sham state $\Phi(\br_1\sigma_1...\br_N\sigma_N; 0)$: $v\xc(\br,t) = v\xc[n; \Psi(0),\Phi(0)](\br,t)$. 
But Eqs.~(\ref{eq:vxcW}) and (\ref{eq:vcT})
depend explicitly on instantaneous many-body quantities such as the xc hole density $\n\xc(\br,\br',t)$ and the 1RDM $\rho_1(\br',\br,t)$, as well as the Kohn-Sham 1RDM $\rho_{1,{\sss S}}(\br',\br,t)$,  and we do not know 
the functional form of these in terms of the density and the initial wavefunctions.  Operating TDDFT via the time-dependent Kohn-Sham equations offers a partial simplification: one may search for explicit functionals of 
the time-evolving Kohn-Sham orbitals, which are themselves implicit functionals of $n(\br,t)$ and $\Phi(0)$, and, by putting more information in the variables that the functional depends on in this way, one hopes that
the functional form itself can be simpler. For example, the Kohn-Sham 1RDM is directly accessible in terms of the instantaneous Kohn-Sham orbitals, $\rho_{1,{\sss S}}(\br',\br,t) = \sum_{i,occ.} \phi_i^*(\br', t)\phi_i(\br,t)$,
and has implicit dependence on $n(\br,t'<t)$ and $\Phi(0)$, including memory. The terms depending on $n\xc$ and $\rho_1$ are however still unknown as functionals of $\phi_i(\br,t)$. 
  
In the next section we propose some preliminary approximations to ${v\xc^W}$ and ${ v\c^T}$ in terms of the instantaneous Kohn-Sham orbitals (orbital functionals) 
and many-body quantities evaluated at initial time (frozen approximations), that can be used in practical propagation schemes~\footnote{The initial many-body quantities are obtained from an interacting many-body computation at the initial time, consistent with the exact functionals being dependent on the initial many-body wavefunction}. We will assume for simplicity that the boundary term in Eq.~(\ref{eq:vcT}) is zero, which is the case for the one-dimensional systems that we will test our approximations on. 

\subsection{Approximations arising from the exact decomposition}
\label{sec:approxs}

\subsubsection{$v\xc\S$}
Here we replace all the quantities from the interacting wavefunction with Kohn-Sham quantities on the RHS of Eq.~(\ref{eq:3Dvxc}), defining a "single-particle" approximation that we denote by $v\xc\S$. 
It is then immediately evident that the kinetic term goes to zero, and such an approximation is an approximation of 
the interaction component $v\xc^W$, making the following replacement in the RHS of Eq.~(\ref{eq:vxcW}):
\ben
v\xc\S: \quad n\xc \to n\xc\S
\label{eq:vxcS}
\een
where the KS xc hole $n\xc\S$ is defined via the KS pair density 
$P\S(\br',\br,t) \equiv N(N-1)\int \vert \Phi(\br',\br,\br_3...\br_N; t) \vert^2 d\br_3..d\br_N 
= n(\br,t)\left(n(\br',t) +n\xc\S(\br,\br',t)\right)$.
 This approximation was first presented in Ref.~\cite{FNRM16}, where the shape and size of $v\xc\S$ was shown to be close 
to $v\xc^W$ for a particular dynamics of a model system. This was field-free dynamics of an initial interacting state prepared in a 50:50 superposition of a ground and first excited-state. Ref.~\cite{FNRM16} 
found that for a wide range of choices of the initial Kohn-Sham state $\Phi(0)$, the approximate potential $v\xc\S$ remained very close to $v\xc^W$ throughout the dynamics. The approximation was also tested for time-resolved electron-atom scattering in Refs.~\cite{SLWM17,LSWM18}, where it captured the approach of the electron to the target far better than the conventional approximations, although, like them, also ultimately severely underestimated the scattering probability. 
The integral in Eq.~(\ref{eq:vxcW}) tends to smooth out details of the xc hole somewhat, making $v\xc^W$ less sensitive to the approximation for the xc hole than $v\c^T$ is to the approximation made for $\Delta\rho_1$ in Eq.~(\ref{eq:vcT}) as we will see shortly. 

From its definition, $v\xc\S$ is an orbital functional, and thus generally is non-adiabatic in that it depends implicitly on the history of the density and on the Kohn-Sham initial state. When a Slater determinant 
is chosen for the Kohn-Sham initial state,  it is highly plausible that $v\xc\S$ reduces to time-dependent exact-exchange (TDEXX). For two electrons this follows directly from the formulae, but for more electrons, 
it is not as straightforward to show~\footnote{The reasoning goes as follows.  TDEXX is defined via the time-dependent optimized effective potential~\cite{UGG95}, which is derived from a stationary principle based 
on an action obtained from evaluating $H - i\partial/\partial t$ on a Slater determinant state, with $H$ being the full interacting Hamiltonian~\cite{UGG95,L98,LHRC17}. If instead the exact interacting wavefunction was used, 
the procedure would yield the exact xc potential (and in general needs to be done on the Keldysh contour~\cite{L98}); TDEXX can be thought of as assuming $\Phi = \Psi$ is a Slater determinant in such a procedure. 
Now the force-balance equation for $\ddot{n}$ which gives rise to Eq.~(\ref{eq:3Dvxc}) also yields the exact xc potential, and $v\xc\S$ results if we make the exact same 
approximation, $\Phi= \Psi$ = Slater determinant in this equation. The latter corresponds to $v\x^{Loc}$ in Ref.\cite{RB09}, which
the authors compare against exchange-only-KLI for some particular dynamics and find the two yield similar results. It is extremely difficult to show directly from the formulae that the two expressions are equal~\cite{RB09b}, but the argument above suggests it is likely that they are.}.
For a more general $\Phi(0)$, $v\xc\S$ goes beyond exact-exchange, and contains a portion of correlation~\cite{FNRM16}. 
In either case, it has spatial- and time- non-local dependence on the density, and dependence on the KS initial state, through the dependence on the KS orbitals.

\subsubsection{$v\xc\S + v\c^T(0)$}
We now address approximations for the kinetic term to be used in conjunction with $v\xc\S$. 
Our first approximation that has a non-zero contribution from the kinetic part of the correlation potential  is very simple: we freeze it to its initial value, that is, on the right-hand-side of Eq.~(\ref{eq:vcT}) 
 \ben
v\c^{T}(0): \quad \Delta \rho_1(t) \to \Delta \rho_{1}(0); \quad  \quad  n(t) \to n(0).
\een
This is of a somewhat similar spirit to the frozen approximations that are popular in strong-field atomic and molecular physics, 
such as the single-active electron approximation \cite{AVSCD08} where it is assumed that only one electron 
is responsible for the dynamics while all the rest are frozen, 
such that if the approximation was self-interaction free, the potential that the dynamic electron sees would be static.
It is also similar to the "instantaneous ground-state" approximation of Ref.~\cite{MRHG14} when any externally applied field is static. 
In addition to the non-adiabatic effects contained in $v\xc\S$ this approximation also is dependent on the initial interacting and KS states through its dependence on $\Delta\rho_1(0)$.

\subsubsection{$v\xc\S + v\c^{T,\Delta\rho_1(0)}$}
To "thaw" the frozen kinetic component somewhat, we consider now the next simplest approximation, which is to freeze only the 1RDM-difference in $v\c^T$, i.e. 
\ben
v\c^{T,\Delta\rho_1(0)}: \quad \Delta \rho_1(t) \to \Delta \rho_{1}(0)
\label{eq:VcTDelta0}
\een
while self-consistently  updating the density in the denominator on the RHS of Eq.~(\ref{eq:vcT}). 
Like $v\xc\S+v\c^T(0)$, the approximation $v\xc\S + v\c^{T,\Delta\rho_1(0)}$ inherits a degree of non-adiabatic density-dependence and Kohn-Sham initial-state dependence through the instantaneous orbital-dependence in $v\xc\S$, and a degree of both Kohn-Sham and interacting initial-state dependence through the kinetic component.

\subsubsection{$v\xc\S + v\c^{T,\rho_1(0)}$}
Noting that $\rho_{1,{\sss S}}$ is accessible during the Kohn-Sham evolution, we define a further frozen approximation where only the interacting 1RDM is fixed to its initial value, while both the Kohn-Sham 1RDM and the density are self-consistently updated during the time-evolution: 
\ben
v\c^{T,\rho_1(0)}: \quad \Delta \rho_1(t) \to \left(\rho_{1}(0) - \rho_{1,S}(t)\right).
\label{eq:VcTrho10}
\een
This approximation freezes the interacting 1RDM while the Kohn-Sham 1RDM may vary greatly.
 
 In the next sections we will discuss the fulfillment of exact conditions and the performance in reproducing some model dynamics for each of the proposed 
 approximations.
 
\section{Exact conditions}
\label{sec:excond}
Exact conditions have been a key instrument guiding the development of physically-inspired and robust gs density functionals that work reliably over a wide range of systems~\cite{PK03,SRP15}. 
Except for the ones arising from the energy-minimization principle which cannot be extrapolated to the time-dependent case, 
many of the gs conditions should be also fulfilled by the 
exact td density functional ~\cite{TDDFTbook2012} and pose a stringent test for approximations. In addition to these, are conditions that are particular to the time-dependent case, and involve aspects of memory-dependence~\cite{M16}.
In this section we analyze the approximations proposed in the 
previous section in the light of the fulfillment of the exact conditions enumerated below. 

(i) Zero Force Theorem (ZFT)

The zero force theorem~\cite{GDP96,OL90,V95b,V95} ensures that the xc potential does not exert a net force,
  \ben
\int  n(\br,t) \nabla v\xc[n](\br,t) d^3 r = 0. 
\label{eq:ZFT}
\een
Since the net force exerted by the Hartree potential vanishes, Eq.~(\ref{eq:ZFT}) ensures that the inter-electron Coulomb interaction doesn't exert
any force on the system
\footnote{In current-density functional theory there is also a zero-torque version of this equation, but it does not hold generally in TDDFT where the Kohn-Sham current might differ from the exact by a rotational component).}.
Linear response (around the gs) applied on Eq.~(\ref{eq:ZFT}) yields a link between the xc kernel $f\xc(\br, \br', \omega)$ and the gs
xc potential $v\xc^0(\br)$ \cite{Vignalechap} that shows that frequency-dependence in a memory-dependent kernel is incompatible with a concurrent local-in-space density-dependence. (The Gross-Kohn functional~\cite{GK85} thus violates ZFT). Violation of ZFT has been shown to lead to numerical instabilities~\cite{MKLR07} due to the system self-exciting. 

(ii) Generalized Translational Invariance (GTI)

Translational invariance requires the wavefunction in an accelerated frame (boost) to transform as,
\ben
|\Psi^b (\br_1...\br_N,t) \rangle
=  \prod^N_{j=1}e^{-i \br_j \cdot\dot{\bb}(t)}|\Psi (\br_1+\bb(t)...\br_N+\bb(t),t)\rangle
\label{eq:boostedwf}
\een
where $\bb(t)$ is the position of the accelerated observer and   
$\bb(0)=\dot{\bb}(0)=0$ such that accelerated and inertial systems coincide at initial time. 
The boosted density transports 'rigidly', 
\ben
n^b(\br,t)=n(\br+\bb(t),t).
\label{eq:rigiddens}
\een
Vignale proved in Ref.~\cite{V95} that 
in order to fulfill GTI the xc potential must transform according to,
\ben
  v\xc^b[n; \Psi(0), \Phi(0)](\br,t) = v\xc[n;\Psi(0), \Phi(0)](\br+\bb(t),t)
  \label{eq:GTI}
\een
A $v\xc$ that fulfills Eq.~(\ref{eq:GTI}) automatically fulfills the ZFT Eq.~(\ref{eq:ZFT}) \cite{V95}.
 A special case of GTI is the harmonic potential theorem (HPT), which states that for a system 
 confined by a harmonic potential and subject to  a uniform time-dependent electric field, the density transforms rigidly following
 Eq.(\ref{eq:rigiddens}) where $\bf{b}(t)$ is the position of the center of mass 
\cite{D94,V95,GDP96,Carstenbook}. (It was shown that Gross-Kohn functional also violates GTI \cite{V95}).

(iii) Memory Condition

The memory condition states that any instant in time can be regarded as the "initial" moment if the wavefunction is known at that time, and the xc potential should be invariant as to which previous time is used in its functional-dependence~\cite{MBW02}, i.e. 
 \ben
 v\xc[n_{t'};\Psi(t'), \Phi(t')] (\vec{r},t)\; \mbox{is independent of}\; t' {\rm for}\; t> t', 
 \label{eq:memory}
 \een
 where $n_{t'}(\br, t) = n(\br,t)$ for  $t>t'$ and is undefined for $t<t'$.
This is a very strict condition on the xc potential, tying together initial-state and history dependences, and has implications also when the initial-state is a ground-state.

 (iv) Self-interaction Free (SI free)
 
 As in the ground-state case, the realization that an electron does not interact with itself yields the condition that for any one-electron system the correlation potential is zero and  exchange cancels the Hartree potential,
  \ben
  v\x^{\rm 1e}({\br,t})= -\int dr' n(\br,t)w(|\br-\br'|), \quad v\c^{\rm 1e}(\br,t)=0.
  \label{eq:si-free}
  \een

  (v) Constant Resonances Condition (CRC)
  
 If a system in an arbitrary superposition state, $\Psi(t)=\sum_n C_n\Psi_n e^{-iE_n t}$,  evolving in a static Hamiltonian, is probed to determine the excitation frequencies, $\omega_i$, then the frequency predicted for a given transition should be independent of the state, i.e. of the $C_n$. 
Within TDDFT this condition requires a subtle cancellation between time-dependence of the xc kernel and the KS response function in a generalized non-equilibrium response formalism~\cite{FLSM15,LFM16}: 
  the exact (non-equilibrium) Hartree-xc kernel must compensate time-dependence in the poles of the KS response function in order to maintain  
   constant $\omega_i$.  
  Given an approximation, this condition is difficult to test analytically; numerically it can be tested by turning off the external field at a given time $\mathcal{T}$ and kicking the system. The frequencies of the ensuing dipole moment should be independent of $\mathcal{T}$ if they satisfy this condition. Further, the fact that different choices of KS initial state give different peak positions is also an indication of CRC violation~\cite{LFM16}.

  \subsection{Table of results}
  \label{sec:excondresults}
  First, let us make some observations regarding the exact components of the xc potential; these are summarized in Table  \ref{tab:excom}. All the proofs may be found in Appendix~\ref{sec:Appendix}.
  The exact interaction and kinetic components, Eqs.~(\ref{eq:vxcW}) and (\ref{eq:vcT}) independently fulfill  ZFT, GTI, memory and are SI free.   We can further break down the kinetic component $v\c^T$ into a part $v_{int}^T$ 
  that depends on the interacting 1RDM and another $v_{\sss S}^{T}$ that depends on 
  the KS 1RDM:
$v_{int}^T$ corresponds to replacing $\Delta \rho_1 \to \rho_1$ and $v_{\sss S}^T$ corresponds to replacing 
$\Delta \rho_1 \to \rho_{1,\sss S}$  in Eq.~(\ref{eq:vcT}). So $v\c^T[n,\Delta \rho_1]  = v_{int}^T[n,\rho_1] - v_{\sss S}^T[n,\rho_{1,S}]$. 
  The two components $v_{int}^T$ and $v_{\sss S}^{T}$  fulfill ZFT independently. GTI is only fulfilled by the difference of the two terms. We  therefore expect a strong violation of GTI when the kinetic component is approximated by a term that asymmetrically approximates the interacting and KS 1RDMs (as in the case of $v\xc\S + v\c^{T,\rho_1(0)}$).
None of the individual components of the exact $v\xc$ satisfy independently the CRC in general; the full sum of these terms is generally required. 

 In Table \ref{tab:approx} we summarize the results for the approximations presented in section \ref{sec:approxs}; the proofs can be found 
 in Appendix~\ref{sec:Appendix} for all conditions except the CRC. The latter is difficult to prove analytically, so we rely on numerical examples, and in all cases we found one example which explicitly showed peak shifting indicative of the violation of this condition.   We separate the analysis of the $v\xc\S$ component from the kinetic components, considering their satisfaction or violation of the exact conditions independently, although in practise we always evolve a kinetic component in conjunction with $v\xc\S$.
 The potential $v\xc\S$ fulfills all exact conditions considered here, except for the CRC, and presents the more robust performance as we shall see in section 
 \ref{sec:runs}, although it does not always give results that are the ones closest to the exact. On the other hand, $v\c^T(0)$ violates all of them except  SI-free. Freezing 
 the difference in the 1RDMs, $v\c^{T, \Delta \rho_1(0)}$  fulfills ZFT and is SI free; it turns out to perform quite well for some of the dynamics when used in combination with $v\xc\S$ as we will see in section~\ref{sec:runs}. However GTI is violated since the system does not
 transform properly under a boost, but the violation is likely weaker than that for $v\c^{T, \rho_1(0)}$: there are large cancellations between  $v_{int}^T$ and $v\s^T$ that cannot occur when
 freezing $\rho_1$ while evolving $\rho_1\S$ (see discussion later). 
 $v\c^{T, \Delta \rho_1(0)}$ also violates the memory condition and the CRC. 
 On the other hand $v\c^{T, \rho_1(0)}$ fulfills ZFT but, as mentioned before, violates 
 GTI strongly due to the asymmetric treatment of the interacting and Kohn-Sham 1RDMs.
 Further, this approximation violates the memory condition and is not SI-free.
In table \ref{tab:approx} we also include the adiabatic local approximation (ALDA) and the adiabatically-exact (AE) approximation 
$v\xc^{AE}$, the later corresponds to propagating with the exact 
ground-state xc potential evaluated on the instantaneous density, $v\xc^{AE}= v\xc^{gs}[n(t)]$. 
$v\xc^{AE}$ fulfills GTI and therefore also ZFT and so does ALDA \cite{D94b,V95,V95b}, but ALDA  is not SI-free. Because of the lack of 
dependence on the initial interacting and KS wavefunctions and on the history of the density, both adiabatic approximations trivially fulfill the memory condition~\cite{MBW02}. 
But both ALDA and AE violate the CRC as was shown in  Refs.~\cite{FLSM15,LFM16} respectively.


   \begin{table}[h!]
  \centering
  \resizebox{0.5\textwidth}{!}{  
    \begin{tabular}{|c|c|c|c|c|c|}\hline
     Exact components  & ZFT & GTI & Memory cond & SI free & CRC
       \\\hline
         { {$v\xc^W$}} 
&  yes & yes & yes & yes & no \\ \hline
  { {$v\c^T=v_{int}^T-v_{S}^T$}} 
&  yes & yes & yes & yes & no \\ \hline
  { {$v_{int}^T$}} 
&  yes & no & yes & no & no \\ \hline
  { {$v_S^T$}} 
&  yes & no & yes & no & no \\ \hline
\end{tabular}
  }
   \caption{Fulfillment of exact conditions by the exact components of the xc potential.}
\label{tab:excom}
\end{table}

 \begin{table}[h!]
  \centering
  \resizebox{0.5\textwidth}{!}{  
    \begin{tabular}{|c|c|c|c|c|c|}\hline
 Approximations     & ZFT & GTI & Memory cond & SI free & CRC
       \\\hline
   { {$v\xc\S$}} 
&  yes & yes & yes & yes & no\\ \hline
  {{$v\c^{T}(0)$}} 

& no & no & no & yes & no \\ \hline
{{$v\c^{T, \Delta \rho_1(0)}$}} 

& yes & no & no & yes & no \\\hline
{{$v\c^{T, \rho_1(0)}$}} 

& yes &  no & no & no & no\\\hline

{{$v\xc^{AE}$}} 

& yes &  yes & yes & yes & no\\\hline
{{$v\xc^{ALDA}$}} 

& yes &  yes & yes & no & no\\\hline

  \end{tabular}
  }
   \caption{ Fulfillment of exact conditions by the proposed approximations to the interaction and kinetic components. Also included are the adiabatically-exact (AE) and ALDA approximations. }
    \label{tab:approx}
  \end{table}
 
 \section{Analysis for non-equilibrium dynamics}
\label{sec:runs}
In this section we analyze the performance of the non-adiabatic approximations introduced in section \ref{sec:approxs} to simulate  
electron dynamics.  We focus on systems driven well beyond linear response from their gs, i.e. non-equilibrium situations,
where we expect memory (non-adiabaticity) 
to be quite relevant.
 In general, the performance of an xc functional approximation will likely depend on the system, the particular dynamics under study, and the observables of interest; as evident from the literature, even the simplest approximation, ALDA, has a performance range of abysmal to accurate enough to be very useful~\cite{M16}. This should be borne in mind when a limited range of studies is done, as here, however at the same time, these studies could be carefully used to indicate expected behavior of the functionals in more general situations. 
 
We choose a numerically exactly-solvable model system for which we can compute the exact evolution of the many-body wavefunction.  
We compare not only the performance of the approximation to simulate the density evolution but also 
the features of the potential as the dynamics evolves. The system is a one-dimensional model of the Helium atom: two soft-Coulomb
interacting electrons in a soft-Coulomb well (atomic units are used throughout),
\ben
\begin{split}
\hat{H} = &\sum_{i=1,2}\left(-\frac{1}{2}\frac{d^2}{dx_i^2} - \frac{2}{\sqrt{x_i^2+1}} + v_{\rm app}(x_i,t)\right)
\\ &+\frac{1}{\sqrt{|x_1-x_2|^2+1}}.
\end{split}
\label{eq:H}
\een
where $v_{\rm app}(x,t)$ is an applied potential, mimicking a laser field. 

All computations were performed using an in-house code.
  The spatial length of the simulation box is 40 a.u. with spacing 0.1 a.u. and has absorbing boundaries; the time-step used was $\leq 0.02$ a.u. 
\subsection{Field-free evolution of a superposition state}
 Our first case study is a superposition state evolving freely, i.e. $v_{\rm app}(x,t)=0$. We prepare the He atom in a 50:50 superposition of the ground and first-excited singlet 
 state and we let it evolve freely,
\ben
\Psi(0) = \frac{1}{\sqrt{2}}\left(\Psi_0 + \Psi_1\right),
\label{eq:Psi5050}
\een
where $\Psi_i=\Psi_i(x_1,x_2)$ denote the many-body eigenstates of Hamiltonian Eq.~(\ref{eq:H}) with $v_{\rm app}(x,t)=0$. The frequency of the dynamics corresponds to the 
energy difference between ground and first excited state, $\omega_0=E_1-E_0 = 0.534$ a.u. (period $T_0 = 2\pi/\omega_0  \approx 12$ a.u).

Our proposed approximate functionals depend on the choice for the KS initial state $\Phi(0)$. Note that this is a separate statement than simply the fact that different initial states yield different dynamics under the same potential: here the potential itself depends on the initial state. The theorems of TDDFT state that any KS initial state may be chosen provided it has the same initial density and initial first time-derivative of the density as the true interacting state. The exact xc potential is different for each choice, but yields the same density-dynamics for each choice. The potentials in an adiabatic approximation are invariant to the choice and yield different density-dynamics for each (and some investigations have been made whether one can judiciously pick an initial Kohn-Sham state for a given physical state in which an adiabatic approximations performs best~\cite{EM12,FNRM16,SLWM17}).  Like the exact xc potential, our approximate potentials are sensitive to the choice of initial KS state, but unlike the exact xc potential, the 
dynamics 
differs greatly depending on this choice. 

We explore the following choices:
\newline
 1. single Slater Determinant (SSD): 
 \ben
\Phi(x_1,x_2,0) =\varphi(x_1)\varphi(x_2)
 \label{eq:SSD}
 \een
 where $\varphi(x) = \sqrt{n(x,0)/2}$, with $n(x,0)$ the initial density of the interacting system,
\newline
2. non-interacting 50:50 superposition:
  \ben
  \Phi^{50:50}(0)=\frac{1}{\sqrt{2}}\left(\Phi_0(x_1,x_2)+\Phi_1(x_1,x_2)\right),
  \label{eq:Phi5050}
  \een
  where $\Phi_0$ is the ground state and $\Phi_1$ is the first non-interacting singlet single excitation 
 \ben 
 \Phi_1(x_1,x_2)=\frac{1}{\sqrt{2}}\left(\varphi_0(x_1)\varphi_1(x_2) + \varphi_1(x_1)\varphi_0(x_2)\right),
 \een
with $\varphi_0$ and $\varphi_1$ the ground and first excited orbitals of a non-interacting potential such that the one-body density of $\Phi^{50:50}(0)$, $n_{\Phi^{50:50}(0)} = \frac{1}{2}\left(3|\varphi_0(x)|^2 + |\varphi_1(x)|^2\right) +\sqrt{2}Re[\varphi_0^*(x)\varphi_1(x)] = n(x,0)$.   

To obtain  $\varphi_0(x)$ and $\varphi_1(x)$ we use an iterative procedure that targets the initial density. The procedure is close to the ones found in Ref.~\cite{TGK08,PNW03}.
  At every iteration the eigenstates $\varphi_i(x)$ are found for a guess of the potential $v\s(x)$ and a density $n(x)$ is obtained from $\Phi_{0}$ or $ \Phi^{50:50}(0)$. Then a correction in the potential is added, proportional to $n(x)-n_{\mathrm{target}}(x)$. Here we actually use $\lambda(n(x)-n_{\mathrm{target}}(x))/(n(x)+\varepsilon)$ where $\lambda$ and $\varepsilon$ are small numerical parameters. The convergence criterium is the difference between the two densities.


We compare the densities and potentials resulting from propagating  these initial KS states under our approximations, with the exact ones.

To find the exact time-dependent potential, a variant of the global fixed point iteration method of \cite{NRL13,RPL15} was used.
While the method of Ref.~\cite{NRL13} has been designed to also have a high accuracy even in the low density regions,
which requires a careful treatment of the boundary conditions and low density regions,
we are not so much focused on low densities and tail regions in this work.
We therefore adopted a more pragmatic approach, introducing a cut-off that flattens the potential in the small-density regions.
The cut-off takes the form $g(x) = \tfrac{1}{2} + \tfrac{1}{2} \tanh[\alpha(n(x)-c)]$
where $c$ is the cut-off and $\alpha$ is a large constant.
For a potential of the form $v(y) = \int^y f(x) dx$ applying the cut-off corresponds to $v(y) = \int^y g(x) f(x) dx$.
Since the method of \cite{NRL13} in each time-step tries to get the right density,
and thus within each time-step to compensate the error in the density caused by for example the cut-off,
it causes an overly oscillatory potential in the low density regions.
We therefore further limited the number of iterations to three to avoid this issue.
As initial guess we used the analytic formulas, Eqs.~(\ref{eq:vxcW}) and (\ref{eq:vcT}), since we also propagate the correlated wave function.
(Though analytically exact, these formulas alone do usually not lead to a stable numerical propagation:
the numerical error created by the multiple derivatives and integrations lead to a density that drifts away from the reference density).

\subsubsection{SSD choice for $\Phi(0)$}
\label{sec:FFa0}
  \begin{figure}
 \includegraphics[width=0.5\textwidth]{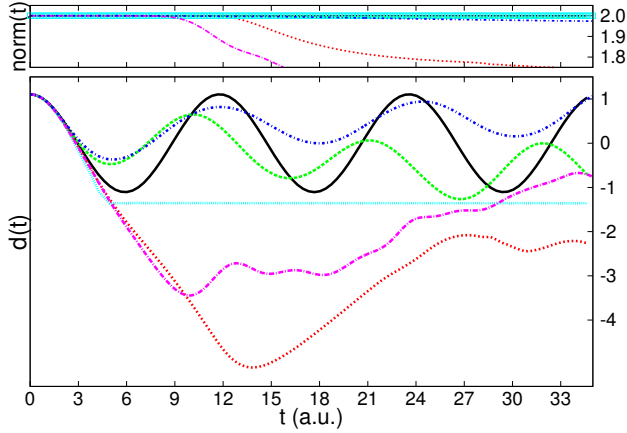}
 \caption{Field-free dynamics of the 50:50 superposition state Eq.~(\ref{eq:Psi5050}) where the approximate TDDFT calculations propagate the SSD choice for initial KS state $\Phi(0)$, Eq.~(\ref{eq:SSD}).
 Norm (upper panel) and dipole moment (lower panel) for exact (black), $v\xc\S=$AEXX (green),  $v\xc\S+ v\c^{T}(0)$ (magenta), $v\xc\S+ v\c^{T,\Delta \rho_1(0)}$ (red), $v\xc\S+ v\c^{T, \rho_1(0)}$ (light blue)  and  $v\xc\ALDA$ (blue).  Time is given in a.u. here and in all following graphs. }
 \label{fig:ffa0dipole}
  \end{figure}



\begin{figure}
      \begin{center}
 \includegraphics[width=0.5\textwidth]{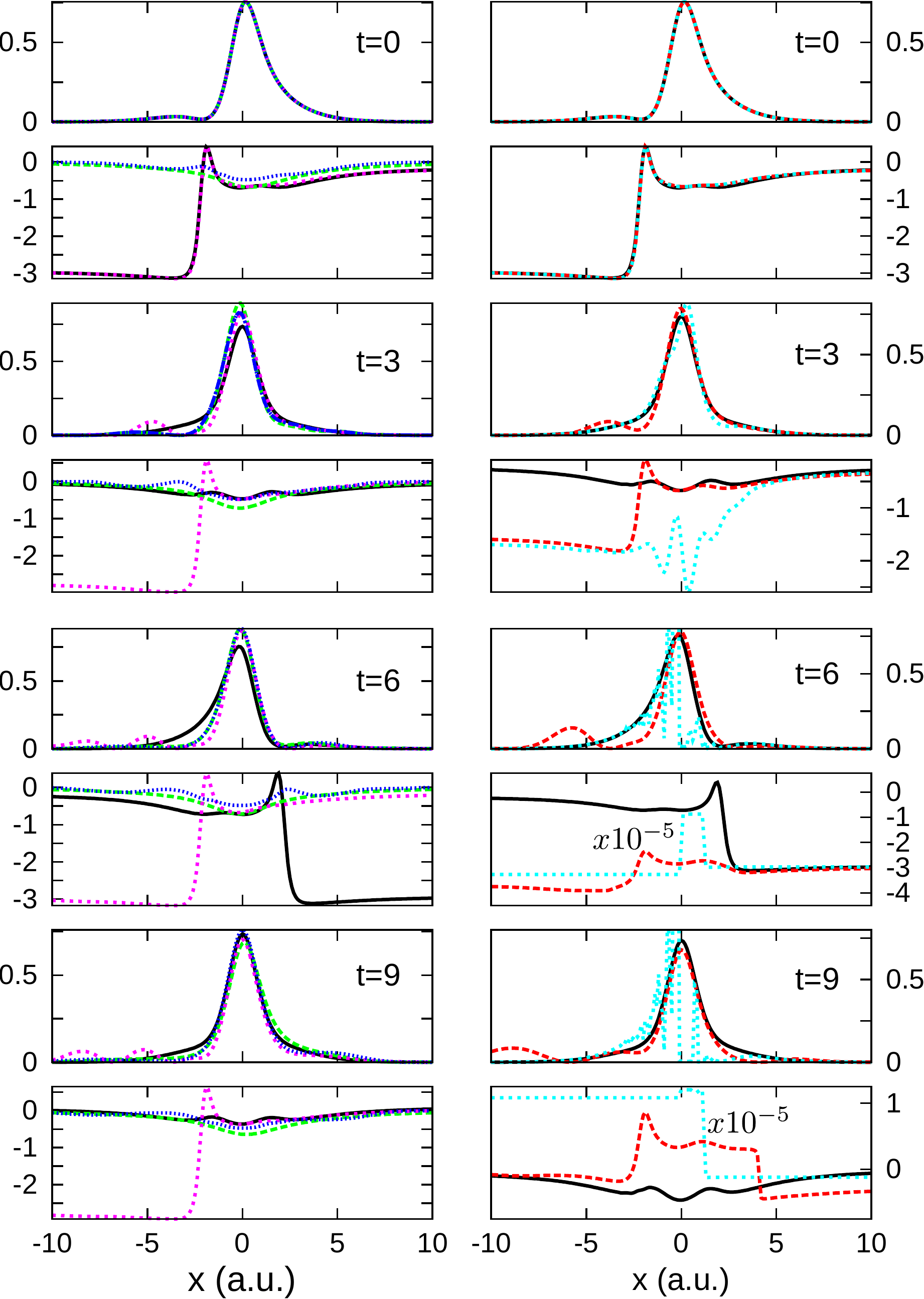}
 \caption{
   Density (upper panel) and $v\xc$ potential (lower panel) for the same dynamics as Fig.\ref{fig:ffa0dipole} shown at initial time, and at three other times indicated within the first cycle. Left panel:  exact (black), $v\xc\S=$AEXX (green), $v\xc\S + v\c^T(0)$ (magenta) and $v\xc\ALDA$ (blue). Right panel: exact (black), $v\xc\S + v\c^{T,\Delta \rho_1(0)}$ (red) and $v\xc\S + v\c^{T,\rho_1(0)}$ (light blue).  The $v\xc\S + v\c^{T,\rho_1(0)}$ potential is scaled by a factor of $10^{-5}$ at $t=6a.u.$ and $t=9a.u.$. 
}
 \label{fig:ff_a0_Pot}
\end{center}
  \end{figure}

In Fig. \ref{fig:ffa0dipole} we show the dipole $d(t) = \int x  n(x,t) dx$ and norm $N(t) = \int  n(x,t) dx$ for 
the field-free evolution of superposition state Eq.~(\ref{eq:Psi5050}) with first KS initial state choice of the previous subsection, the SSD Eq.~(\ref{eq:SSD}).
The exact dynamics is shown in black. Propagating using $v\xc\S$ (in green) which, for two electrons and SSD choice, corresponds to adiabatic exact-exchange (AEXX) yields a dipole that appears to oscillate with more than one frequency, 
giving an envelope to the oscillations, and neither the dominant frequency nor the amplitude of oscillations are captured well. The ALDA propagation (blue) is also shown, and it approximates the frequency of the oscillation a little better than AEXX, but the amplitude is poor and again there is  a beating over long time. 
The propagations that include the various approximate $v\c^T$ improve over the traditional approximations for only a very short time but soon deviate dramatically with qualitatively wrong behavior even before half a cycle of the evolution. 
We will now discuss what goes wrong with each of these, with the help of the density and potential time-snapshots within the first optical cycle shown in Fig.~\ref{fig:ff_a0_Pot}.

We first note that when a SSD is chosen for the KS system, with one  doubly-occupied orbital, the features of the exact $v\xc$ (in black) have the periodicity of the density dynamics: 
initially the density has a minimum on the left side coincident with a large dynamical step and peak in $v\c^T$, which decreases and disappears as the density 
becomes symmetric around $t=3$a.u., before building up again on the other side, such that at half-cycle ($t\approx 6$ a.u.) $v\xc$ and the density are a mirror images of their initial values.  

Consider now the potentials of the traditional approximations,  $v\xc\S =$AEXX  and ALDA  (in green and blue in Figs.~\ref{fig:ffa0dipole}-\ref{fig:ff_a0_Pot}). Although both the AEXX and ALDA dipoles are close to the exact for $t<3$a.u. the shape of the density at this time has 
already started to deviate. The $v\xc\S$ potential is generally a very good approximation to the exact interaction part $v\xc^W$ of the xc potential~\cite{FNRM16} but it
completely misses the step and peak features present in the exact $v\c^T$,  simply smoothly cradling the density as it evolves, and ALDA also does not capture these structures. 
These missing structures are important to get the details of the dynamics right, including the period, as evident in the plots.

Consider now propagation with $v\xc\S + v\c^{T}(0)$ (magenta curves in Figs.~\ref{fig:ffa0dipole}-\ref{fig:ff_a0_Pot}). As $v\c^T(0)$ is static, the large initial step 
in the kinetic correlation part of the potential remains for all time, and as a consequence the density leaks to the left  and is unable to slosh back to the right (see magenta density in left panel Fig.~\ref{fig:ff_a0_Pot}). 
The density eventually gets absorbed by the boundary around $t\approx 30$a.u. as can be seen in the evolution of the norm in the upper panel of Fig.~\ref{fig:ffa0dipole}.

  
We turn now to propagation under $v\xc\S + v\c^{T,\Delta \rho_1(0)}$, where again the dipole swings far too much to the left (Fig. \ref{fig:ffa0dipole}), 
resulting in a large unphysical absorption evident in the norm in the top panel. 
The reason for this becomes evident from the right panel of Fig.~\ref{fig:ff_a0_Pot}:  Freezing $\Delta \rho_1$ in $v\c^T$  results in a  step that, 
although varying in size, is always on the left, pulling the density consistently over to the left. 

\begin{figure}
 \includegraphics[width=0.5\textwidth]{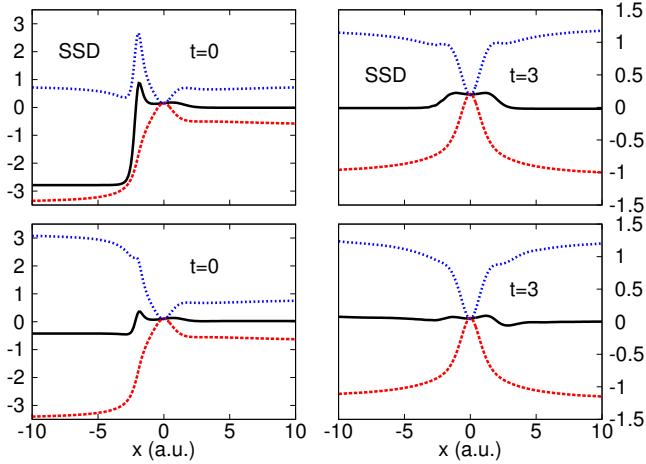}
 \caption{
   Exact components $v_{int}^T$ (in red) and $-v_S^T$ (in blue) of the exact kinetic component $v\c^T=v_{int}^T -v_S^T$ (in black)
   for field-free dynamics evolution of superposition state Eq.~(\ref{eq:Psi5050}) at initial time and at about $1/4$ cycle (t = 3 a.u.). Top  panel:  SSD choice Eq.~(\ref{eq:SSD}) for KS initial state. Lower panel: Superposition choice Eq.~(\ref{eq:Phi5050}) for the KS initial state.}
 \label{fig:stepcomp}
\end{figure}

On the other hand, propagation under  $v\xc\S + v\c^{T,\rho_1(0)}$ yields a dipole that displays a most peculiar flattening after half a cycle (light blue in Figs.~\ref{fig:ffa0dipole} and~\ref{fig:ff_a0_Pot}). The freezing of the interacting 1RDM while keeping the KS 1RDM dynamical has drastic consequences, as can be seen from first recalling the exact $v\c^T$: Both the exact $v_{int}^T$ and $v_S^T$
   display large dynamical steps that tend to counteract each other, yielding a smaller step in the exact $v\c^T$, at times even canceling each other. (see Fig.~\ref{fig:stepcomp} and  movie 1 in supplementary material).  Freezing only the interacting 1RDM as in $v\c^{T,\rho_1(0)}$ leads to   the development of a disproportionately large step at times, resulting in instabilities and the poor dynamics as shown here. 

\begin{figure}
 \includegraphics[width=0.5\textwidth]{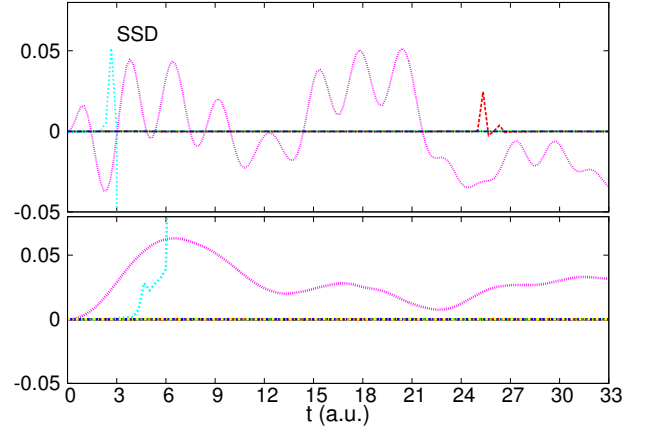}
 \caption{Violation of ZFT: The left-hand-side of Eq.~(\ref{eq:ZFT}) in field-free evolution of superposition state Eq.~(\ref{eq:Psi5050}), in propagation under $v\xc\S$ (green),  $v\xc\S+ v\c^{T}(0)$ (magenta), $v\xc\S+ v\c^{T,\Delta \rho_1(0)}$ (red), $v\xc\S+ v\c^{T, \rho_1(0)}$ (light blue)  and  $v\xc\ALDA$ (blue). Upper panel:  SSD choice Eq.~(\ref{eq:SSD}) for KS initial state. Lower panel: Superposition choice Eq.~(\ref{eq:Phi5050}) for the KS initial state.}
 \label{fig:ZFT}
\end{figure}

Satisfaction of the ZFT is demonstrated in Fig.~\ref{fig:ZFT} which plots the left-hand-side of Eq.~(\ref{eq:ZFT}) for the different approximations. From Table II, all approximations except for $v\xc\S + v\c^{T}(0)$ are expected to satisfy the ZFT, and give zero. The deviation of $v\xc\S+ v\c^{T, \rho_1(0)}$ is a numerical error related to the very large steps that appear in this approximation as just discussed. The spike in $v\xc\S+ v\c^{T,\Delta \rho_1(0)}$  at around 25 a.u. is a numerical artifact caused by the step in the potential at that time being very large and sharp. Notice the relatively large violation of the ZFT for $v\xc\S + v\c^{T}(0)$ in this  dynamics.

 We conclude that except for quite short times, the new approximations do not improve the dynamics of a field-free superposition state when the initial KS state is chosen to be a SSD;  in fact the traditional approximations ALDA and AEXX (coinciding with $v\xc\S$ in this case)  are more reliable, although inaccurate.
  
One might be tempted to argue that these results have limited relevance, since starting the KS system in an SSD when the interacting state is a 50:50 superposition state would be a bad choice, making a challenging job for approximate functionals to do well. However, this could well be the situation reached when the system began in the ground-state, with some field applied that brought the interacting state to a 50:50 superposition state, at which point the field is turned off. Then the natural choice for the initial KS state is the KS ground-state, and we are stuck with a SSD for the whole evolution.

\subsubsection {Superposition choice for $\Phi(0)$}
\label{sec:FFa1}
\begin{figure}
 \includegraphics[width=0.5\textwidth]{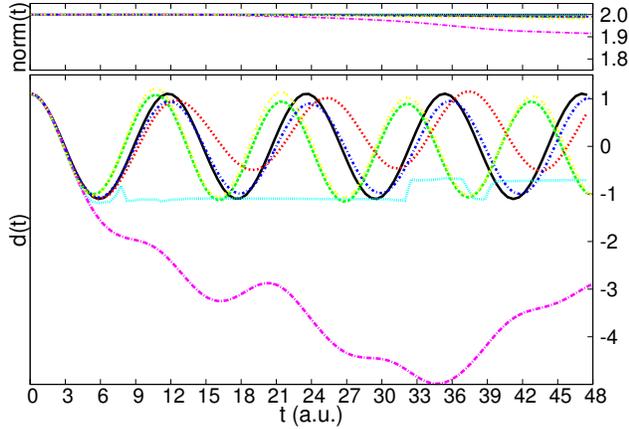}
 \caption{
 Time-dependent dipole moment (lower panel) and norm (top panel) for field-free dynamics of 50:50 superposition state Eq.~(\ref{eq:Psi5050}). Initial KS state $\Phi(0)$ is chosen as in Eq.~(\ref{eq:Phi5050}).  Shown are results of propagation under the exact (black), $v\xc\S$ (green),   $v\xc\S+ v\c^{T}(0)$ (magenta), 
 $v\xc\S+ v\c^{T,\Delta \rho_1(0)}$ (red), $v\xc\S+ v\c^{T, \rho_1(0)}$ (light blue), AEXX$=-\frac{v\H}{2}$ (yellow) and ALDA (blue).
 }
 \label{fig:a1dipole}
\end{figure}

Fig.~\ref{fig:a1dipole} shows the field-free superposition state dynamics for the different approximations when the KS initial state 
is chosen to be  Eq.~(\ref{eq:Phi5050}).
Considering first the traditional approximations ALDA and AEXX $=-v\H/2$, we observe a great improvement in their performance with this choice of KS initial state, compared to the Slater determinant choice. In fact ALDA is remarkably good: choosing a KS initial state structure that is close to that of the true system seems to  override the errors from the ground-state xc assumption inherent in the adiabatic nature of ALDA, its self-interaction error, and its locality in space.  It is the configuration of the initial KS state that is key: if instead we used LDA orbitals but still within the first-excited Slater determinant configuration of Eq.~(\ref{eq:Phi5050}), the dynamics is similar although not as accurate (see shortly).  In fact ALDA outperforms the spatially non-local AEXX, suggesting the importance of an adequate accounting of correlation for this example, with even the local-density ground-state correlation approximation doing a good job here. We will see that freezing the kinetic component of the correlation in various ways in our new approximations, together with the correlation contained in $v\xc\S$, do not provide a sufficiently good correlation component to compete with the simple ALDA for more than short to intermediate times. 

 
Turning then to the new approximations, we first note that, unlike the SSD choice, none of the approximations  lead to the unphysical situation of the density reaching the boundaries and being absorbed, at least  until much later (for $v\xc\S + v\c^{T}(0)$), as shown by the norm. Two of the approximations, namely $v\xc\S$ (which gives dynamics close to AEXX) and $v\xc\S+v\c^{T,\Delta\rho_1(0)}$ do a reasonable job however none of them beats the remarkable performance of ALDA for the dipole dynamics as noted above. Inclusion of $v\c^{T,\Delta \rho_1(0)}$ tends to yield a worse amplitude but a better dominant frequency than $v\xc\S$. Performing a Fourier transform of the dipole yields the frequencies, $\omega=0.515 \pm 0.006$ a.u. for $v\xc\S + v\c^{T,\Delta \rho_1(0)}$ as compared with $\omega= 0.584 \pm 0.006$ a.u. for $v\xc\S$ while the exact frequency is $\omega_0=0.534$ a.u, best approximated by ALDA's frequency of $\omega = 0.527 \pm 0.006$a.u.

\begin{figure}
   \includegraphics[width=0.5\textwidth]{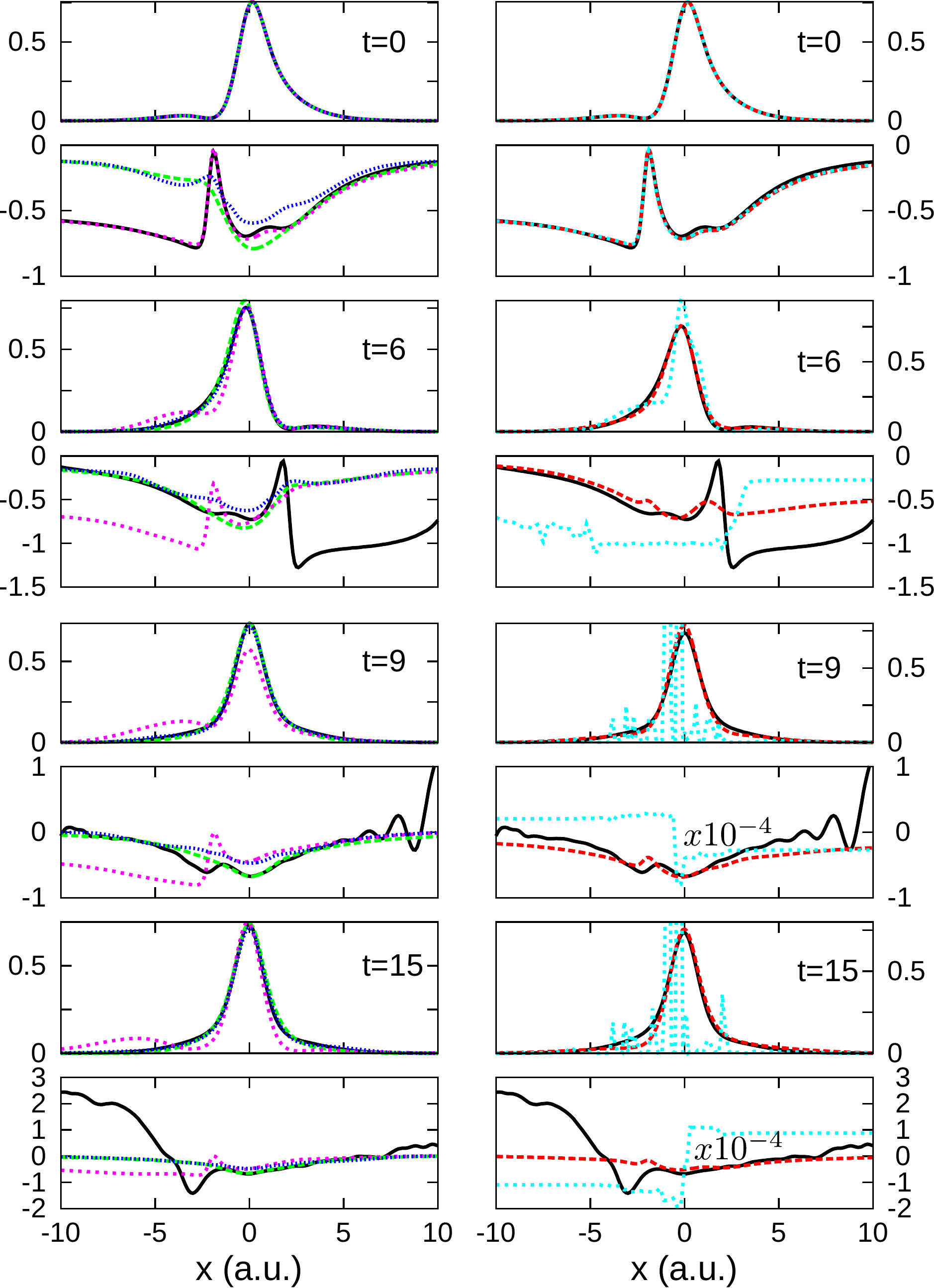}
\caption{
   Time snap-shots of the density (upper panels) and $v\xc$ potential (lower panels) for the dynamics shown in Fig.\ref{fig:a1dipole} for:  exact (black), $v\xc\S$ (green), $v\xc\S + v\c^T(0)$ (magenta),  $v\xc\ALDA$ (blue), $v\xc\S +v\c^{T,\Delta\rho_1(0)}$ (red), and $v\xc\S+ v\c^{T,\rho_1(0)}$ (light blue). The latter has been scaled by $10^{-4}$ for $t = 9$au and $t = 15$au. 
 }
 \label{fig:ff_a1_Pot}
 \end{figure}

These two new approximations, $v\xc\S$ and $v\xc\S +v\c^{T,\Delta\rho_1(0)}$  yield a good density evolution,
as can be seen in Fig.~\ref{fig:ff_a1_Pot}  in green and red respectively, where they can be compared with the exact density and potential (black).   The exact  $v\c^T$ for this choice of KS initial state is initially much smaller than for the SSD choice (c.f. Fig.~\ref{fig:ff_a0_Pot}), and $v\xc\S$, which, as we noted before is a very good approximation to $v\xc^W$,  then plays a relatively more important role than in the SSD case. We note that $v\xc\S$ in this case is {\it not} equivalent to TDEXX: it contains some correlation due to the KS state configuration, and has memory-dependence in that it is an implicit functional of the density at earlier times and of the KS initial state. Including $v\c^{T,\Delta\rho_1(0)}$ now partially captures the  step and peak features of the exact potential and the step "switches sides" unlike for the SSD case. Notice that from  approximately $t=6$a.u. onwards  the exact $v\xc$ is no longer periodic while the density is (unlike the case in the previous section where a SSD is used). The dynamical features of the exact $v\c^T$ become more and more complex as the system evolves
and at later times they become eventually as large as the ones observed for the SSD choice (see also Ref.~\cite{FNRM16}); the movie 2 in supplementary materials displays the densities and xc potentials from exact propagation, ALDA, $v\xc\S$, and $v\xc\S +v\c^{T,\Delta\rho_1(0)}$.
 
 Turning now to the other two approximations, $v\xc\S + v\c^T(0)$ and $v\xc\S+ v\c^{T,\rho_1(0)}$, the performances are poor, although the release of density then absorbed by the boundaries is less compared to the SSD case. This is because the step in the initial kinetic components is less, so even if frozen as in $v\xc\S + v\c^T(0)$, it results in less density moving out to the left. Propagation with $v\xc\S+ v\c^{T,\rho_1(0)}$ suffers from the same problem as in the SSD choice: the uneven treatment of the interacting and KS density matrices results in a huge step in the potential that prevents the density 
 from sloshing back and forth as it should  (right panel of Fig.~\ref{fig:ff_a1_Pot}).

\begin{figure}[h]
 \includegraphics[width=0.45\textwidth, height = 0.45\textwidth]{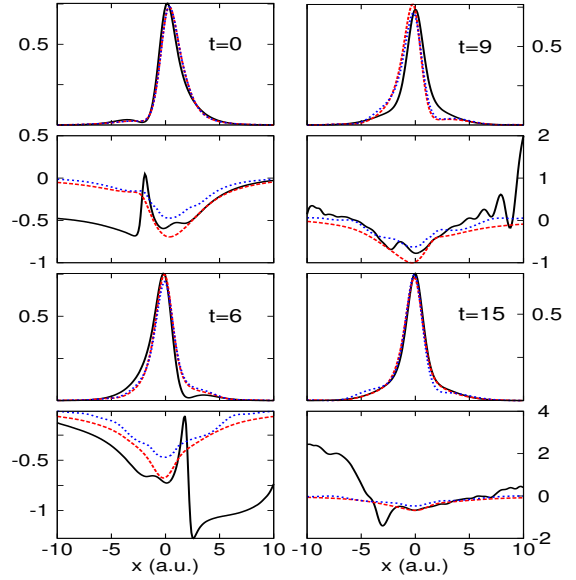}
 \caption{Time snap-shots of the density (upper panel) and $v\xc$ potential (lower panel) for field-free dynamics starting in 50:50 superposition state Eq.~(\ref{eq:Psi5050}) and using KS initial state of the form  Eq.~(\ref{eq:Phi5050}) but with the gs and first excited LDA orbitals.
 Propagation with  $v\xc\S + v\c^{T,\Delta \rho_1(0)}$ (red) and $v\xc\ALDA$ (blue). In black the exact dynamics.}
 \label{fig:ff_a1_delta0_alda}
 \end{figure}

Finally, since typically orbitals that reproduce the exact density are not easily obtained, we show in Fig.~\ref{fig:ff_a1_delta0_alda} the results of propagating a state of the form Eq.~(\ref{eq:Phi5050}) where the LDA ground and first-excited orbitals are inserted into the KS states. We show the best of our new approximations, along with ALDA.  We notice that the density-dynamics is generally significantly worsened for both propagation with ALDA and $v\xc\S + v\c^{T,\Delta \rho_1(0)}$ compared to when initial orbitals that reproduce the exact initial density are used. 
 
\subsection{Arbitrary field starting in the ground state}
\label{sec:arblaser}
\begin{figure}[ht]
 \includegraphics[width=0.5\textwidth]{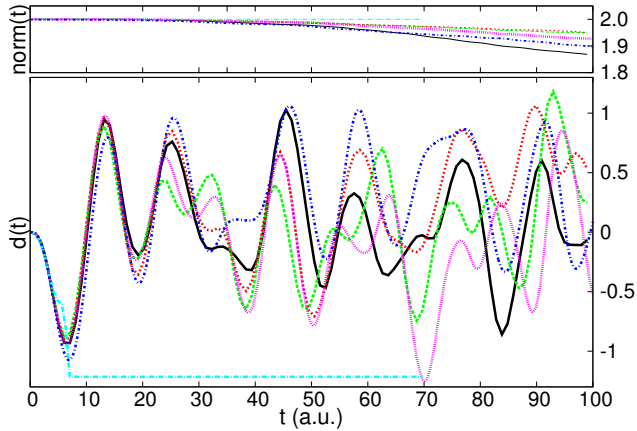}
 \caption{Dynamics of the laser-driven gs. Norm (upper panel) and dipole (lower panel) for exact (black), $v\xc\S$ =AEXX (green),  $v\xc\S+ v\c^{T,\Delta \rho_1(0)}$ (red), $v\xc\S+ v\c^T(0)$ (magenta) and $v\xc\S+ v\c^{T,\rho_1(0)}$ (light blue) and
   $v\xc \ALDA$(blue).}
\label{fig:0_1_0_4_dipoles}
\end{figure}

Next we study the dynamics of system Eq.~(\ref{eq:H}) initially in the gs and driven by a non-resonant, fairly strong, laser field, $v_{\rm app}(x,t) = {\cal E}(t)x$, where ${\cal E}(t)=0.1\sin(0.4 t)$. Since we start in the gs the natural choice is a SSD for the KS initial state, and this is the only choice we consider here.
  In Fig.~\ref{fig:0_1_0_4_dipoles}, the dipoles and norm for the exact and approximate KS evolutions, beginning in the initially exact KS orbital $\varphi(0)=\sqrt{\frac{n_{0}}{2}}$ are shown. 
  
\begin{figure}
 \includegraphics[width=0.5\textwidth]{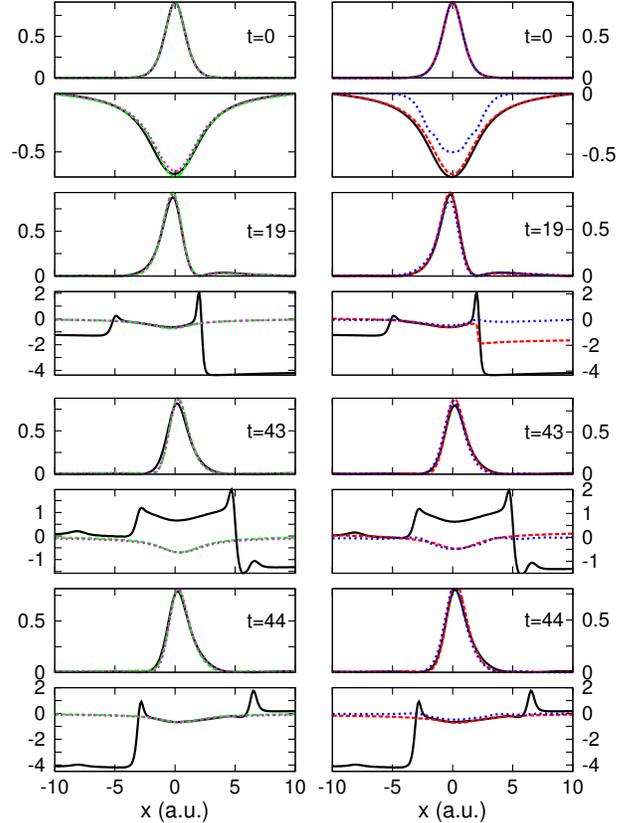}
 \caption{Density (upper panels) and $v\xc$ potential (lower panels) for the laser-driven gs. Left panel: exact (black), $v\xc\S$ =AEXX (green) and $v\xc\S+ v\c^T(0)$ (magenta). Right panel: exact (black) , $v\xc\S+ v\c^{T,\Delta \rho_1(0)}$ (red) and  $v\xc\ALDA$ (blue). }
 \label{fig:laser_pot}
  \end{figure}

The propagation with $v\xc\S$=AEXX alone (in green) is excellent till almost $t=12$ a.u., after which it begins to significantly deviate from the exact. The other traditional approximation ALDA (in blue) is not as good as AEXX in this case for the first few cycles but stays closer to the exact than AEXX at later times. Fig.~\ref{fig:laser_pot} shows that in fact both these  approximations and the new ones, except for  $v\xc\S + v\c^{T,\rho_1(0)}$ which is not shown (see below), capture the structure of the density and the potential quite well in the central region; the noticeable deviations of the dipole moments of the approximations in Fig.~\ref{fig:0_1_0_4_dipoles} arise from errors in the density further out where the density itself is not very large, but these errors get enhanced by the multiplication by $x$ in the calculation of the dipole moment (which, additionally, highlights asymmetric structures in the density).

Adding the correlation potential frozen to its initial value, $v\xc\S + v\c^T(0)$ (in magenta)  does not  improve the results  over $v\xc\S$ (except  at short times) as can be seen by comparing the green and  magenta curves in Fig.~\ref{fig:0_1_0_4_dipoles} and in the density time snapshots of Fig. \ref{fig:laser_pot}. In this case the system starts off weakly correlated, and the exact potential is a smooth well initially, but as the laser field kicks in and begins to drive the interacting system far away from approximately a single Slater determinant, correlation increases, as is reflected in the increased size of the structures in the exact $v\c^T$ evident in Fig.~\ref{fig:laser_pot}; these structures appear in $v\c^T$ rather than in $v\xc^W$ which remains quite smooth. As $v\xc\S + v\c^T(0)$ freezes the $v\c^T$ to its initial value, it cannot capture these. 

The approximation $v\c^{T,\Delta \rho_1(0)}$ does include  partial step structures, although not always quite in phase with the exact (see also the movie 3 in the supplementary material).  
However, these do not apparently have a significant effect on the ensuing dynamics, neither in the region where the density is appreciable, nor in the dipole moment that weighs more the low-density regions further out, and the dipole moment is in fact worse than the traditional approximations except at short times. One aspect of the largest of these step structures for this dynamics, is that they can change very rapidly, for example the large step switches sides from $t = 43$ a.u. to 44 a.u.
 Also, at $t\approx 60$ a.u. the exact dynamics begins to show some absorption (see upper panel Fig.~\ref{fig:0_1_0_4_dipoles}) which is  underestimated in $v\xc\S + v\c^{T,\Delta \rho_1(0)}$. This leads to some discrepancy in the dipole moments, but the density in the central region is very well reproduced still (see lower panel in red in Fig.~\ref{fig:laser_pot}).

  The propagation under  $v\xc\S + v\c^{T,\rho_1(0)}$
  develops very large features near
the boundaries (no compensation of the large step in $v\c\S$, as in the discussion in the previous section) that progressively approach the central region, growing in size, and by $t\approx 5$ a.u. the step has become so large and sharp, that the propagation  results in a freezing of the density on the left (in light blue in Fig.~\ref{fig:0_1_0_4_dipoles}) .  


\begin{figure}
 \includegraphics[width=0.45\textwidth, height = 0.45\textwidth]{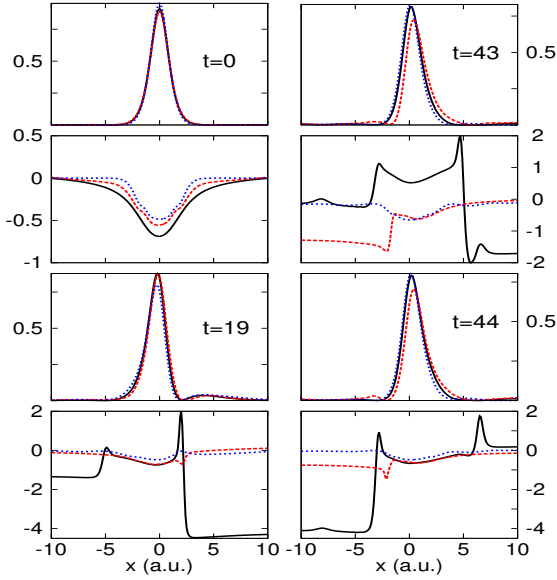}
 \caption{Time snapshots of the density (upper panels) and $v\xc$ (lower panels) for laser-driven dynamics starting in the LDA ground-state: propagation under exact (black), ALDA (blue), $v\xc\S +v\c^{T,\Delta\rho_1(0)}$ (red).}
\label{fig:laser_lda_orbs}
 \end{figure}

Finally, in practice we rarely have access to the exact initial KS orbitals, and approximate DFT orbitals are used for the initial state. In Fig.~\ref{fig:laser_lda_orbs} LDA orbitals
are propagated using $v\xc\S + v\c^{T,\Delta \rho_1(0)}$  and ALDA, showing a noticeable deviation from the results when the exact initial orbitals are used, especially, in this case for the propagation under $v\xc\S + v\c^{T,\Delta \rho_1(0)}$.

\section{Conclusions}
\label{sec:concs}
Building approximations based on the exact expression for the time-dependent xc potential, Eq.~(\ref{eq:3Dvxc}), allows us to immediately break free of the adiabatic approximation in TDDFT. The expression requires approximations for the time-evolving xc hole and for the 1RDM of the interacting system, quantities that are not accessible in a KS calculation. In this work, we have proposed a number of practical approximations to these terms, evaluated their satisfaction of some known exact constraints in TDDFT, and explored their performance on a few model systems. The approximations have memory-dependence, in particular initial-state dependence, and how well they perform in turn depends on the choice of KS state in which to begin the propagation. The functionals can be viewed as orbital-functionals, however they do not require TDOEP in their operation, and have an explicit dependence on the initial interacting many-body state and the initial choice of KS state.

Two of these newly proposed approximations stand out. One is $v\xc\S$ in which all quantities in the exact expression for $v\xc$ are replaced by their KS counterparts. This approximation has been explored in recent literature where it was found to be an excellent approximation to the interaction component $v\xc^W$ of the exact potential~\cite{FNRM16,SLWM17} and to yield a significant improvement in dynamics of an electron approaching a target in scattering compared to ALDA and AEXX~\cite{SLWM17,LSWM18} albeit ultimately failing to actually scatter even qualitatively. In the examples studied here, it provides a reasonable approximation, although ALDA was in fact generally better. A dynamical accounting of correlation beyond what is present in $v\xc\S$ is required. The approximation satisfies the most exact conditions out of the new approximations considered, including ZFT, GTI, SI-free, and the memory condition. 

The second approximation that worked reasonably well was $v\xc\S + v\c^{T,\Delta\rho_1(0)}$, where the kinetic component is approximated by freezing the difference in the 1RDMs to the initial value. In the examples considered here, it did not perform particularly better than the traditional approximations; a better accounting of correlation is required. The approximation satisfies the ZFT and is SI-free, but violates the other exact conditions we considered. A key factor in determining the accuracy of the performance of any approximation, is the choice of the KS initial state; the field-free propagation of the superposition state demonstrated clearly the vast improvement of all approximations once the KS initial state was chosen with a configuration close to that of the interacting state.

It is clear that better approximations are needed for the kinetic term, that involve a dynamical approximation to the 1RDM of the interacting system, and such developments are underway~\cite{LM18}. This term is more challenging as it has much more structure and a stronger non-adiabatic dependence than the interaction term~\cite{LFSEM14}. How well these approximations work for realistic many-electron three-dimensional systems remain to be seen. For robust performance, we expect that certainly the approximation should satisfy the ZFT;  satisfaction of GTI, SI-free, memory-condition, and CRC will likely improve its accuracy. As in the ground-state case, it is not always clear which exact conditions are most important for predicting a functional approximation's success. The approach pointed to in this work, based on approximations to the 1RDM-difference and xc hole appearing in Eq.~(\ref{eq:3Dvxc}), provides a starting point to develop such practical memory-dependent approximations. 

  {\it Acknowledgments:}
Financial support from the US National Science Foundation
CHE-1566197 (NTM) and the Department of Energy, Office
of Basic Energy Sciences, Division of Chemical Sciences,
Geosciences and Biosciences under Award DE-SC0015344 (JIF and LL)
are also gratefully acknowledged. JIF acknowledges CONICET Argentina. SEBN acknowledges financial support from the European Research Council (ERC-2015-AdG-694097)
  and European Union's H2020  programme under GA no. 676580 (NOMAD).

\section{Appendix}
\label{sec:Appendix}
\subsection{Zero Force Theorem Satisfaction}
\label{sec:ZFT}
\subsubsection{Exact components: $v\xc^W$, $v\c^T$, $v_{int}^T$, and $v_S^T$}
Here we prove that $v\xc^W$, $v\c^T$, $v_{int}^T$, and $v_S^T$ independently satisfy the ZFT, Eq.~(\ref{eq:ZFT}). 

Beginning with $v\xc^W$, we write $n\xc(\br,\br',t)$ in terms of the pair density, so that
\ben
\begin{split}
\int n(\br,t) {\bf \nabla} v\xc^W[n](\br,t) d^3r  = \\
\int \int  n(\br,t)   \left(\frac{P(\br',\br,t)}{n(\br,t)} - n(\br',t)\right){\bf \nabla} w(|\br'-\br |)  d^3r d^3r'  
\end{split}
\label{eq:IntnvxcW}
\een
Interchanging $\br\leftrightarrow \br'$ in Eq.~(\ref{eq:IntnvxcW}) we get the same but with opposite sign since 
the integral is antisymmetric due to the gradient, 
 $ \nabla_{\br'} w(| \br'-\br|) =  -\nabla_{\br} w(| \br'-\br|)$.  Thus 
 the only possible solution for Eq.~(\ref{eq:IntnvxcW}) is zero.
 
 Because the exact $v\xc$ satisfies ZFT and $v\xc^W$ does independently, then $v\c^T = v\xc  - v\xc^W$ must also independently satisfy the ZFT. This can also be shown explicitly. In fact, even the exact interacting and KS contributions to $v\c^T$ independently satisfy the ZFT, as we will now show (and hence this also shows explicitly that $v\c^T$ independently satisfies it). 
 
  Consider the force contribution from the interacting part, $v_{int}^T$,
\bea
\begin{split}
& \int  n(\br,t){\bf \nabla} v_{int}^T (\br,t)d^3r = 
\\& \frac{1}{4}\int d^3r \left({\bf \nabla}' - {\bf \nabla} \right)\left(\nabla^2- \nabla'^2\right)\left(\rho_1(\br',\br,t) \right)\vert_{\br'=\br} 
\\ & =\frac{N}{2} Re \int d^3r d^3r_2..d^3r_N \left(\nabla\Psi^* \nabla^2\Psi - \Psi^*\nabla^3\Psi\right)
\\ & = \frac{N}{4} \int d^3r d^3 r_2..d^3r_N \nabla \left(4\vert \nabla\Psi\vert^2 - \nabla^2\vert\Psi\vert^2\right)
\label{eq:ZFTvintT}
\end{split}
\eea
where arguments $(\br,\br_2...\br_N)$ are understood for the wavefunctions, and ${\bf \nabla}$ is the gradient with respect to $\br$. To obtain the third line, the definition of the 1RDM in terms of the wavefunction is inserted, and we have noted that ${\bf \nabla}' = {\bf \nabla}_{r'}$ acts only on $\Psi^*(\br',\br_2..\br_N)$ while ${\bf \nabla} = {\bf \nabla}_r$ acts only on $\Psi(\br,\br_2..\br_N)$.   Noting that the integrand in the final line is in fact a total gradient, we see that upon performing the integral over $\br$, the integral vanishes for finite or periodic systems where the integrand vanishes at infinity or is identical at either end of the periodic system. Hence, the ZFT is satisfied by $v_{int}^T$. 
An identical argument can be made using KS wavefunctions, to show that $v_S^T$ also satisfies the ZFT. 

We conclude that both  $\int  n(\br,t){\bf \nabla} v_{int}^T (\br,t)d^3r$  and $\int  n(\br,t){\bf \nabla} v_S^T (\br,t)d^3r$ vanish independently, and therefore $v\c^T$ fulfills  ZFT independently of $v\xc^W$.

\subsubsection{Approximations: $v\xc\S,v\c^T(0),v\c^{T,\Delta\rho_1(0)},v\c^{T,\rho_1(0)}$}
That $v\xc\S$ satisfies the ZFT follows from the same argument as applied to $v\xc^W$ in the previous section, noting that the KS pair density $P\s(\br,\br',t)$ obeys the same symmetry relations as for the true pair density. 

Turning to the frozen kinetic components, consider first $v\c^T(0)$, for which
\bea
\nonumber
\int n(\br,t)\nabla v\c^{T}(\br,0) d^3r = \\
\int d^3 r \frac{n(\br,t)}{n(\br,0)}\int d^3r' (\nabla' - \nabla)(\nabla^2 - \nabla'^2)\Delta\rho_1(\br',\br,0)\vert_{\br' = \br}
\eea
At the initial time, the right-hand-side is zero, as follows from the previous section, but at later times, due to the ratio of the densities factor, the arguments of the previous section cannot be made, and  
there is no guarantee that the integral will be zero. In fact it is generally non-zero, and therefore $v\c^T(0)$ violates the ZFT. This is also evident in the numerics as can be seen in Fig.~\ref{fig:ZFT} and
the violation leads to numerical instabilities from "self-excitation"of the system~\cite{MKLR07}.

Fortunately, the other two frozen approximations do satisfy the ZFT. By keeping the density in the denominator of Eq.~(\ref{eq:vcT}) dynamical and equal to the evolving density, the density factor in the zero force 
expression $\int n(\br,t){\bf \nabla} v\c^{T,\Delta\rho_1(0)}d^3 r$ is exactly cancelled and the argument follows closely to that of Eq.~(\ref{eq:ZFTvintT}) and subsequent discussion in the previous subsection.
Likewise, the proof there that the two components of $v\c^T$ independently satisfy the ZFT, ensures that $v\c^{T,\Delta\rho_1(0)}$ and $v\c^{T,\rho_1(0)}$ also satisfy it. 

\subsection{Generalized Translational Invariance Satisfaction}
\label{sec:GTIproof}

\subsubsection{Exact components: $v\xc^W$, $v\c^T$, $v_{int}^T$, and $v_S^T$}

From the expression, Eq.~(\ref{eq:GTI}), for the boosted wavefunction discussed in section \ref{sec:excond} it follows that the xc hole transforms as
$n\xc^\bb(\br_1,\br_2,t) = n\xc(\br_1 + \bb(t),\br_2 + \bb(t),t)$ under a boost. Given the invariance of $w(|\br_1-\br_2|)$ under such a
transformation  we conclude that the $v\xc^{W}$ component fulfills GTI,
\ben 
v\xc^{W,\bb}[n\xc](\br,t)= v\xc^W[n\xc](\br + \bb,t).
\label{eq:GTIvxcW}
\een

To study the translational invariance of the kinetic correlation potential we again separate it into
interacting and KS contributions,  $ \nabla v\c^T=\nabla v_{int}^T- \nabla v\s^T$, and study the behavior of each term separately under the boost. That is, 
\ben
\nabla v_{int}^{T,\bb}(\br,t) = \frac{1}{4 n^\bb(\br,t)}\left(\nabla' - \nabla\right)\left(\nabla^2 - \nabla'^2\right)\rho_1^\bb(\br',\br,t)\vert_{\br' = \br}
\een
where
\ben
\rho_1^\bb(\br',\br,t) = e^{-i \dot{\bb}(t)\cdot(\br - \br')}\rho_1(\br'+\bb(t),\br+\bb(t),t)
\een
Working through the derivatives eventually leads to 
\bea
\nonumber
\nabla v_{int}^{T,\bb}(\br,t) &=&
\nabla v_{int}^{T}(\br^\bb(t),t)  - \frac{\left(\dot{\bb}(t)\cdot\nabla\right)\bj(\br^\bb(t),t)}{n(\br^\bb(t),t)}\\
&+&  \frac{\dot{n}(\br^\bb(t),t)}{n(\br^\bb(t),t)} \dot{\bb}^2(t) - \frac{\bb(t)\cdot\nabla n(\br^\bb(t),t)}{n(\br^\bb(t),t)}
\label{eq:vint_b}
\eea
where $\br^\bb(t) \equiv \br + \bb(t)$. 

Eq.~(\ref{eq:vint_b}) shows that $\nabla v_{int}^T$ does not satisfy GTI by itself, since the last three terms on the right are not generally zero. 
However, once we subtract the corresponding equation for $\nabla v\s^T$, noting that the KS system has the same density as the interacting system, we find
\ben
\nabla v\c^{T,\bb}(\br,t) = \nabla v\c^T(\br^\bb(t),t) - \frac{\left(\dot{\bb}(t)\cdot\nabla\right)\left(\bj(\br,t) - \bj\s(\br,t)\right)\vert_{\br = \br^\bb(t)}}{n(\br^\bb(t),t)}
\een
where $\bj\s(\br,t)$ is the current-density of the KS system. 
In 1D, the second term vanishes as the KS and interacting currents are the same, but in 3D they are generally not the same.
Yet, we know that the exact full $v\xc$ satisfies GTI, and also that the exact $v\xc^W$ does, and therefore $v\c^T = v\xc - v\xc^W$ must also. The resolution lies in the additional term $\frac{\nabla \times a(\br,t)}{n(\br,t)}$ in Eq.~(\ref{eq:vcT}) that we assumed zero throughout most of this paper: in 3D this term is not generally zero, and it plays a crucial role here in restoring GTI. 
 
In summary, although $\nabla v_{int}^T$ and $\nabla v\s^T$ separately are in general not translationally invariant, their difference, 
\ben 
v\c^{T,\bb}[n,\Delta \rho_1](\br,t)= v\c^T[n, \Delta \rho_1](\br + \bb(t),t).
\label{eq:GTIvcT}
\een
does satisfy GTI. 

\subsubsection{Approximations: $v\xc\S,v\c^T(0),v\c^{T,\Delta\rho_1(0)},v\c^{T,\rho_1(0)}$}
The same argument used in previous section to prove the exact $v\xc^W$ fulfills GTI can be carried over for the single-particle approximation 
$v\xc\S$ since the KS xc hole transforms properly under a boost, namely $n\xc^{S,\bb}(\br_1,\br_2,t) = n\xc\S(\br_1 + \bb(t),\br_2 + \bb(t),t)$,
\ben 
v\xc^{S,\bb}[n\xc\S](\br,t)= v\xc\S[n\xc\S](\br + \bb(t),t).
\label{eq:GTIvxcW}
\een

However none of the frozen approximations to $v\c^T$ satisfy the GTI. 
In $v\c^T(0)$ the  entire kinetic potential remains constant in time, so does not transport rigidly
under a boost. Likewise it can be seen that freezing $\Delta\rho_1$ in $v\c^{T,\Delta\rho_1(0)}$ or $\rho_1$ alone in $v\c^{T,\rho_1(0)}$ will not transform correctly under a boost.

\subsection{Memory Condition Satisfaction}
Adiabatic xc functionals trivially fulfill the memory condition, Eq.~(\ref{eq:memory}) because of the lack of dependence on the initial interacting and KS states $\Psi(0)$ and $\Phi(0)$. 
When we introduce partial memory as in the approximations proposed here, this condition becomes susceptible to being broken.

\subsubsection{Exact components: $v\xc^W$, $v\c^T$, $v_{int}^T$, and $v_S^T$}

The memory condition is fulfilled by the exact interacting component, $v\xc^W$, 
\ben
v\xc^W[n_{t'},\Psi(t'),\Phi(t')](\br t) = v\xc^W[n\xc[\Psi(t)](t)](\br,t)
\label{eq:MCvxcW}
\een
which is independent of $t'<t$, 
because it depends explicitly on the exact xc hole, $n\xc(\br,\br',t)$, at time $t$, which, although an implicit functional of the history of the density and the initial interacting state,  it is an explicit and instantaneous functional of $\Psi(t)$, from which it is obtained directly.
Likewise, the memory condition is fulfilled by the two components of the exact correlation potential independently: $v_{int}^T$ explicitly depends on the instantaneous interacting 1RDM, an implicit functional of the history of the density and the initial interacting state, but is obtained explicitly from the instantaneous interacting state,
 \ben
 v_{int}^T[n_{t'}, \Psi(t'), \Phi(t')](\br, t) = v_{int}^T[\rho_1[\Psi(t)](t)](\br,t)
  \een
  and so again is independent of $t'<t$. 
 An analogous argument can be made for $v\s^T$, and so 
 the kinetic component of the correlation  potential satisfies the memory condition,
  \ben
  v\c^T[\Delta  \rho_1[n_{t'}, \Psi(t'), \Phi(t')](t)](\br,t)  
  \een
 because it is independent of $t'<t$.
Note that the boundary term $(\nabla \times a(\br,t))/n(\br,t)$ does not affect this analysis, since it depends on instantaneous quantities.

\subsubsection{Approximations: $v\xc\S,v\c^T(0),v\c^{T,\Delta\rho_1(0)},v\c^{T,\rho_1(0)}$}
In the case of $v\xc\S$ the proof follows a similar path as for the exact interaction component Eq.~(\ref{eq:MCvxcW}). The potential $v\xc\S$  is an explicit functional of the KS xc hole $n\xc\S(t)$ which is obtained explicitly from the
 instantaneous KS wavefunction $\Phi(t)$, hence
\ben
v\xc\S[n_{t'},\Psi(t'),\Phi(t')](\br t) = v\xc\S[n\xc\S[\Phi(t)](t)](\br,t)
\een
is independent of $t' <t$. 

In the case of the frozen approximations to the kinetic component, we first note that these approximations are defined with freezing parts of the kinetic component to their values at the time that is considered the initial time, i.e. the time $t'$, in $v\xc[n_{t'}, \Psi(t'), \Phi(t')](\br t)$. 
Therefore, freezing the entire kinetic potential as in  $v\c^T(0)$  means freezing it to $v\c^T(t')$. This clearly is $t'$-dependent, and so violates the memory condition. 
 The approximation $v\c^{T,\Delta\rho_1(0)}(t)$ also violates the condition, since  
 \ben
 \nabla v\xc^{T,\Delta\rho_1(0)}[n_{t'}, \Psi(t'),\Phi(t')](t) = \frac{\mathcal{D}\Delta\rho_{1}(\br',\br,t')|_{\br'=\br}}{4n(\br,t)} 
 \een
 where $\mathcal{D} = (\nabla' - \nabla)\left(\nabla^2- \nabla'^2\right)$
 will not in general be independent of $t'$. 
 Likewise $v\c^{T,\rho_1(0)}(t)$ will violate it in general, due to the $t'$-dependence of the interacting 1RDM in this approximation.

\subsection{Self-interaction}
\subsubsection{Exact components: $v\xc^W$, $v\c^T$, $v_{int}^T$, and $v_S^T$}
The exact interaction component cancels the Hartree component for one electron as it should,
\ben
\nabla v\xc^W(\br,t)= -\int d\br' n(\br',t)\nabla w(|\br-\br'|)=-v\H(\br,t)
\een
since $n\xc(\br',\br,t)=-n(\br',t)$ for $N=1$. 
The two exact components of the kinetic potential are not independently SI-free,
\ben
v_{int}^T\neq 0 \quad {\rm and} \quad v_{S}^T\neq 0 \quad {\rm for} \quad N=1
\een
but it is the subtraction of the two what gives the exact kinetic correlation component, which properly vanishes for one electron, 
 \ben
 v\c^T(\br,t)=0 \quad (N=1)
 \een
since for $N=1$, $\Psi(t)=\Phi(t)$ and thus $\rho_1(t)=\rho_{1,S}(t)$.

\subsubsection{Approximations: $v\xc\S,v\c^T(0),v\c^{T,\Delta\rho_1(0)},v\c^{T,\rho_1(0)}$}
That $v\xc\S$ is SI-free follows directly the argument above for  $v\xc^W$, since for one electron $\Psi(t)=\Phi(t)$ , the pair-density is zero, and $n\xc\S(\br,\br',t) = n\xc(\br,\br',t)=-n(\br',t)$.
The approximations $v\c^{T}(0)$ and $v\c^{T,\Delta \rho_1(0)}$ are both SI-free, because for one electron $\rho_1(0) = \rho_1\S(0)$, so 
\ben
v\c^{T}(0)(\br,t)=0 \quad {\rm and} \quad v\c^{T,\Delta \rho_1(0)}(\br,t)=0 \quad (N=1)
\een
On the other hand, for $v\c^{T,\rho_1(0)}$ the uneven treatment of the interacting and KS 1RDMs results, in general, in a SI error,
    \ben
    v\c^{T,\rho_1(0)}(\br,t)\neq 0 \quad  t>0 \quad (N=1).
    \een

\addcontentsline{toc}{section}{References}
\bibliography{./ref_exp}

\end{document}